\documentclass[twocolumn,showpacs,prl,superscriptaddress]{revtex4}
\usepackage{amsmath}   
\usepackage{graphicx}
\usepackage{bm}
\usepackage{color}
\usepackage{amsfonts}

\begin{document}
\title{Universal features of dynamic heterogeneity in supercooled liquids}
\author{Elijah Flenner}
\affiliation{Department of Chemistry, Colorado State University, Fort Collins, CO 80523}
\author{Hannah Staley}
\affiliation{Department of Physics, Colorado State University, Fort Collins, CO 80523}
\author{Grzegorz Szamel}
\affiliation{Department of Chemistry, Colorado State University, Fort Collins, CO 80523}

\date{\today}

\begin{abstract}
A few years ago it was showed that \emph{some} systems that have very similar local structure, as quantified by 
the pair correlation function, exhibit vastly different slowing down upon supercooling 
[L. Berthier and G. Tarjus, Phys. Rev. Lett. \textbf{103}, 170601 (2009); 
U.R. Pedersen, T.B. Schr{\o}der and J.C. Dyre, Phys. Rev. Lett. \textbf{105}, 157801 (2010)].
Recently, a more subtle structural quantity, the so-called ``point-to-set'' length, was 
found to reliably correlate with the average dynamics [G.M. Hocky, T.E. Markland and 
D.R. Reichman, Phys. Rev. Lett. \textbf{108}, 225506 (2012)]. Here we use computer simulations to 
examine the behavior of fluctuations around the average dynamics, \textit{i.e.}, dynamic heterogeneity. 
We study five model glass-forming liquids: three model liquids used in previous 
works and two additional model liquids with finite range interactions. Some of these systems have very
similar local structure but vastly different dynamics. We show that for all these systems
the spatial extent and the anisotropy of dynamic heterogeneity correlate very well with the average dynamics. 
\end{abstract}

\pacs{61.20.Lc, 61.20.Ja, 64.70.Q-}

\maketitle
Upon supercooling, universal phenomena are observed in seemingly unrelated glass-forming
systems. Similarly, glass transition theories predict universal relationships between
different static and dynamic quantities. Some of the relationships predicted by the theories are 
difficult to verify experimentally but they can be tested in computer simulations. These tests 
can help to differentiate between different theories. 
Due to the large computational 
resources required, simulations often examine one relatively simple model system.
However, this does not establish that the relationships 
between different static or dynamic quantities are truly universal.
Here, we examine universal features of dynamic heterogeneity, \textit{i.e.}, fluctuations around
the average dynamics. 

Our study is inspired by a re-evaluation of the standard van der Waals picture \cite{CWA1983} 
of the liquid state in the context
of supercooled liquids' dynamics. Within the standard picture, 
the liquid's local structure, as quantified by the pair correlation function,
is primarily determined by the repulsive part of the interparticle potential. Importantly,
it was believed (admittedly, with somewhat limited simulational \cite{KushickBerne1973,YoungAndersen2005} 
and theoretical \cite{Bembenek2000} support) that the 
local structure, and thus the repulsive part of the potential, also determines the liquid's dynamics.
Therefore, it was surprising when Berthier and Tarjus \cite{Berthier2009} showed that two standard model liquids,
which differ only by the presence of the attractive part of the potential and have very
similar local structure, exhibit vastly different viscous slowing down upon approaching the glass transition.
Shortly after this work, Pedersen \textit{et al.}\  \cite{Pedersen2010} complicated the picture 
by finding a system with a purely repulsive potential, the same local structure, and 
the same dynamics as the model liquid with both repulsive and attractive interactions. 

More recently, Hocky \textit{et al.} \cite{Hocky2012} investigated a different, more subtle, static quantity, the 
so-called ``point-to-set'' length scale \cite{BouchaudBiroli2004} 
in the systems considered by Berthier and Tarjus, and Pedersen  \textit{et al.} 
Hocky \textit{et al.} found that the point-to-set length
can have different values for systems with very similar local structure, but it 
correlates very well with the average dynamics and shows universal features for
all the systems studied. 

We present results of an extensive computer simulation study that tests 
the universality of fluctuations around the average dynamics, \textit{i.e.}, 
dynamic heterogeneity. First, we investigate two standard quantities used to characterize dynamic heterogeneity,
the four-point susceptibility, which measures the overall strength of the
heterogeneity, and the dynamic correlation length, which measures the spatial extent of the heterogeneity. 
In addition, we calculate 
quantities that are sensitive to the
anisotropy of dynamic heterogeneity. Investigation of the latter quantities has been prompted by recent 
experiments of Zhang \textit{et al.}\ \cite{Zhang2011}, who studied two glassy colloidal systems that differed by the 
presence of an attractive part of the effective colloid-colloid potential. They found profound dependence 
of the shape of the clusters of fast particles on the presence of the attractions. 

For large enough supercooling, we find that all quantitative characteristics of dynamic heterogeneity for 
all systems investigated have the same dependence on the relaxation time that characterizes the 
average dynamics. 

We divide the systems we studied into two groups. The systems in the first group (which were also 
studied by Hocky \textit{et al.}) are derived from
the Kob-Andersen binary Lennard-Jones mixture \cite{Kob1994,Kob1995a,Kob1995b}. 
We simulated the standard Kob-Andersen mixture (KA), the
Weeks-Chandler-Andersen (WCA) truncation \cite{Weeks1971, Chandler1983} of the standard mixture, and a system 
with an inverse power law (IPL) potential that was designed by Pedersen \textit{et al.} \cite{Pedersen2010}. 
All three systems have similar pair-correlation functions at the same temperature. However, only the KA and IPL 
mixtures exhibit the same temperature dependence of the relaxation time 
\cite{Berthier2009,Pedersen2010}. We studied dynamic properties of these systems as a function of temperature 
at a fixed volume using Newtonian dynamics.

The second group consists of two 50:50 mixtures of spherical particles with the same size ratio. The first system is 
a hard sphere (HARD) system, where the particle positions 
are updated using Monte-Carlo dynamics with local moves \cite{Flenner2011,Berthier2007}. 
The second system is 
a repulsive harmonic sphere (HARM) system \cite{BerthierWitten2009}. 
The HARM system was studied using Newtonian and Brownian dynamics.  
The control parameter for the
hard sphere system is the volume fraction, while it is the temperature for the harmonic spheres. 

The details of the simulations and the reduced units which we use to present our results 
are given in the Supplemental Material \cite{supplement}. 

To find a correlation between dynamic heterogeneity and the average dynamics in systems with different potentials, 
different control parameters and different underlying microscopic dynamics we need to define a rescaled 
relaxation time. To this end we use a hallmark property of supercooled liquids: violation of the Stokes-Einstein 
relation. In the normal liquid state the Stokes-Einstein relation holds and the self-diffusion coefficient is 
inversely proportional to the relaxation time, 
$D \sim \tau_\alpha^{-1}$. The violation of this relation in supercooled liquids 
is frequently associated with the appearance of dynamic heterogeneity \cite{BerthierDH,Ediger2000}.
We define a rescaled relaxation time in such a way that all systems we study deviate from
the Stokes-Einstein relation at the same rescaled relaxation time. 

We calculate the self-diffusion coefficient $D$ from the  mean-square-displacement, 
$D = \lim_{t \rightarrow \infty} (6tN)^{-1} \left< \sum_n |\mathbf{r}_n(t) - \mathbf{r}_n(0) |^2 \right>$.
We define the alpha relaxation time $\tau_\alpha$ in terms of the self-intermediate scattering function
$F_s(k;t)$ using the standard relation $F_s(k_{0};\tau_\alpha) = e^{-1}$, where
$k_{0}$ is chosen to be around the first peak of the static structure factor $S(k)$.  
For the KA, WCA, and IPL systems
$k_{0} = 7.2$, and for the HARM and HARD systems $k_{0}=6.1$. 
\begin{figure}
\includegraphics[width=3.2in]{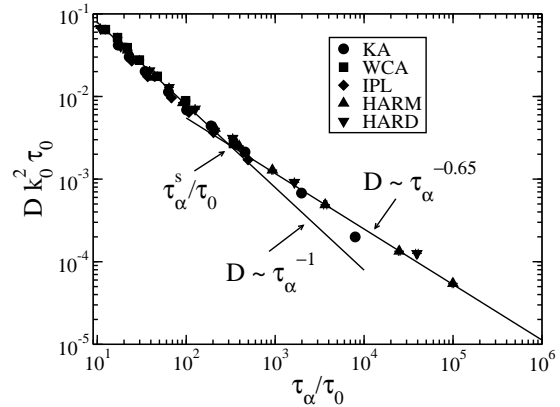}
\caption{\label{Dtau}Rescaled self-diffusion coefficient $D k_{0}^2 \tau_0$ plotted versus 
rescaled relaxation time $\tau_\alpha/\tau_0$ for all the systems investigated. The lines are
fits of the HARM data to $D \sim \tau_\alpha^{-1}$ for $T \ge 12$ and to a fractional Stokes
Einstein relation $D \sim \tau_\alpha^{-\lambda}$ for $T \le 7$. 
The rescaled relaxation time $\tau_{\alpha}^s/\tau_0$ is when these two fits are 
equal.}
\end{figure}

To find the rescaling of the relaxation time, we used the HARM system as a reference. 
For the remaining systems, we rescaled the relaxation time $\tau_\alpha$ by a constant $\tau_0$ 
so that these systems deviate from the  Stokes-Einstein relation at the same $\tau_\alpha/\tau_0$. This procedure
results in $\tau_0 \approx 1/15$ for the KA, WCA, and IPL systems, and $\tau_0 \approx 70$ 
for the hard-sphere system. Somewhat unexpectedly, we found that by plotting $D k_0^2 \tau_0$ as a function
of $\tau_\alpha/\tau_0$ we obtain a reasonable collapse of all the data, see Fig.~\ref{Dtau}.  

We note that a crossover time scale (defined by crossing point of 
two power-law relations showed in Fig.~\ref{Dtau}) is equal to $\tau_\alpha^s = 303 \tau_0$. 
This time scale corresponds to a temperature $T_s$
(or volume fraction $\phi_s$) located between 
the onset of glassy dynamics and the mode-coupling transition temperature
(see Supplemental Material \cite{supplement} for more details).

To obtain the four-point susceptibility and the dynamic correlation length we start with 
an often studied four-point structure factor defined in terms of overlap functions pertaining to
individual particles, 
\begin{equation}
\label{s4over}
S_4^{\mathrm{ov}}(q;t) = \frac{1}{N} \left<
\sum_{n,m} w_n(a;t) w_m(a;t) 
e^{i \mathbf{q} \cdot [\mathbf{r}_n(0) - \mathbf{r}_m(0)]} 
\right>.
\end{equation}
Here $w_n(a;t)$ is the overlap function, $w_n(a;t) = \Theta[a-|\mathbf{r}_n(t) - \mathbf{r}_n(0)|]$, where 
$\Theta(x)$ is Heaviside's step function.  
$S_4^{\mathrm{ov}}(q;t)$ is the structure factor of the particles that move less than a distance
$a$ over a time $t$, and it is used to characterize the size of clusters of slow particles.
We calculate this structure factor at time $\tau_\alpha^{\mathrm{ov}}$,
which is defined in terms of the average overlap function 
$F_s^{\mathrm{ov}}(\tau_\alpha^{\mathrm{ov}}) = N^{-1} \left< \sum_N w_n(a;\tau_\alpha^{\mathrm{ov}}) \right> 
= e^{-1}$. We choose
$a$ such that $\tau_\alpha^{\mathrm{ov}}$ is close to the relaxation time $\tau_\alpha$ defined 
in terms of the self-intermediate scattering function.
For the KA, WCA, and IPL systems $a=0.25$ and for the HARM and HARD systems
$a=0.3$ (note that these choices make the product $k_0 a$ approximately the same for all systems investigated). 
We use the previously described procedure \cite{Flenner2010,Flenner2011} to calculate the 
four-point susceptibility $\chi^\mathrm{ov}_4$ and the dynamic correlation length $\xi_4^{\mathrm{ov}}$.

\begin{figure}
\includegraphics[width=3.2in]{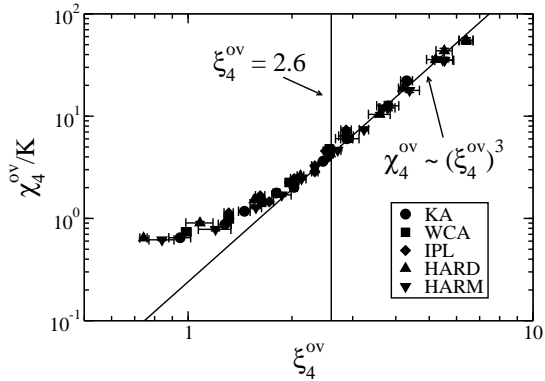}
\caption{\label{rchixi}A rescaled dynamic susceptibility $\chi_4^{\mathrm{ov}}/K$ versus $\xi_4^{\mathrm{ov}}$.
Here $K$ is a system dependent scaling constant. $K$ is the same for the KA, WCA, and IPL systems. 
}
\end{figure}

\begin{figure}
\includegraphics[width=3.2in]{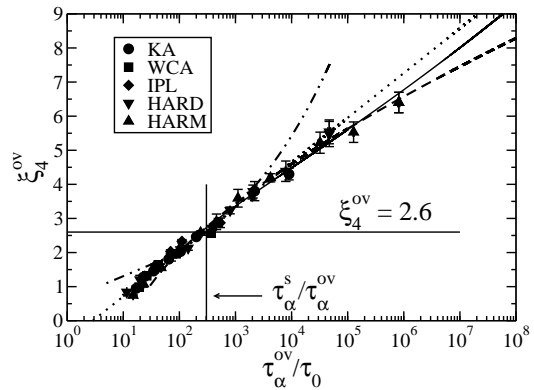}
\caption{\label{rtauxi}The four-point correlation length $\xi_4^{\mathrm{ov}}$ versus the 
rescaled relaxation time $\tau_\alpha^{\mathrm{ov}}/\tau_0$. The lines are fits to different theoretical 
functions. The dash-dotted line is a mode-coupling-like fit 
$\xi_4^{\mathrm{ov}} \sim (\tau_\alpha^{\mathrm{ov}})^{1/z}$ where $z=4.8$. 
The next two fits are inspired by RFOT theory. 
The dotted line is a fit to $\xi_4^{\mathrm{ov}} \sim \ln(\tau_\alpha^{\mathrm{ov}})$.
For longer relaxation times the data are better described by 
$\xi_4^{\mathrm{ov}} \sim \ln(\tau_\alpha^{\mathrm{ov}})^{3/2}$ which is showed as the dashed line. 
The solid line is a fit to the facilitation 
prediction of $\ln(\xi_4^{\mathrm{ov}}) = A \sqrt{\ln(B \tau_\alpha^{\mathrm{ov}})} + C$.
We also indicated 
the relaxation time where Stokes-Einstein violation begins (solid vertical line), $\tau_\alpha^s$, and 
the correlation length where the relationship $\chi_4^{\mathrm{ov}} \sim (\xi_4^{\mathrm{ov}})^3$ 
begins (solid horizontal line).
}
\end{figure}

First, we investigate
the relationship between these two quantities. In Fig.~\ref{rchixi}
we show $\chi_4^{\mathrm{ov}}/K$ plotted versus $\xi_4^{\mathrm{ov}}$. Here 
$K$ is a system dependent scaling constant. We found that $K$ 
is the same for the KA, WCA, and IPL systems. 
For $\xi_4^{\mathrm{ov}} > 2.6$ we find that 
$\chi_4^{\mathrm{ov}}$ grows as $(\xi_4^{\mathrm{ov}})^3$ for all
systems investigated. 
We note that $\xi_4^{\mathrm{ov}} = 2.6$ 
when the system's relaxation time is $\tau_\alpha^s$, Fig.~\ref{rtauxi}. 
We recall that the Random-First-Order Theory (RFOT) 
approach predicts compact dynamically correlated regions for temperatures below
the mode-coupling transition temperature \cite{Stevenson2006}. 
We find $\chi_4^{\mathrm{ov}}\propto (\xi_4^{\mathrm{ov}})^3$, which indicates compact clusters of slow particles,
starting from the crossover temperature $T_s$ (or volume fraction $\phi_s$).

We now examine the correlation between the dynamic correlation length $\xi_4^{\mathrm{ov}}$ calculated at
$\tau_\alpha^{\mathrm{ov}}$ and $\tau_\alpha^{\mathrm{ov}}/\tau_0$. Note that 
to define a rescaled time scale we use the 
values of $\tau_0$ which were determined before by analyzing the relation between $D$ and $\tau_\alpha$. 
This is justified since the temperature (or volume fraction) dependence of $\tau_\alpha^{\mathrm{ov}}$
and $\tau_\alpha$ is very similar.  
We note that the results for all systems investigated collapse onto the same curve when plotted
as $\xi_4^{\mathrm{ov}}$ versus $\tau_\alpha^{\mathrm{ov}}/\tau_0$, Fig.~\ref{rtauxi}. 
While we anticipated having to rescale $\xi_4^{\mathrm{ov}}$, this does not seem necessary for the systems
studied. 
We conclude that the spatial extent of dynamic
heterogeneity correlates very well with the average dynamics  when the average dynamics is rescaled
relative to the point at which the Stokes-Einstein relation is violated. 

We compare our results to three theoretical scenarios. The relationships between 
the dynamic correlation length and the relaxation time obtained from these scenarios are showed as 
lines in Fig.~\ref{rtauxi}. 
We find that a power law relationship between $\xi_4^{\mathrm{ov}}$ and $\tau_\alpha^{\mathrm{ov}}$ 
(dash-dotted line) obtained from a mode-coupling-like approach \cite{Biroli2004,Biroli2006,Szamel2008} 
is a poor description of the data for more than about a decade of slowing down. 
Next, we find that a logarithmic relationship $\xi_4^{\mathrm{ov}} \sim \ln(\tau_\alpha^{\mathrm{ov}})^{1/\zeta}$, 
inspired by an Adam-Gibbs like \cite{Adam1965} or 
a Random-First-Order Transition (RFOT) theory \cite{Kirkpatrick1989,Lubchenko2007}, describes well 
the initial slowing down with $\zeta = 1$ (dotted line) but at longer relaxation times $\zeta = 2/3$ 
(dashed line)
provides a better fit. 
Finally, the relation inspired by the facilitation picture, 
$\ln(\xi_4) = A \sqrt{\ln(B \tau_\alpha^{\mathrm{ov}}/\tau_0)} + C$ (solid line) \cite{Keys2011},
is also compatible with the data. In principle, a more detailed analysis of the existing data 
(including independent estimates of various theoretical parameters) may be able to distinguish between the 
latter two approaches. We note, however, that the theoretical scenarios are nearly indistinguishable 
over a large range of $\xi_4$ versus $\tau_\alpha^{\mathrm{ov}}/\tau_0$. 
The most direct comparison would be enabled by extending the
range of the available (rescaled) relaxation times by some two orders of magnitude. 

Fig.~\ref{rchixi} indicates a change in the spatial organization of dynamic heterogeneity. 
This fact, together with experimental finding of Zhang \textit{et al.}\ \cite{Zhang2011}, prompted us to 
examine in some detail the shape of dynamic heterogeneity. 
To this end we study a four-point structure factor defined in terms
of microscopic self-intermediate scattering functions pertaining to different particles, 
\begin{equation}
S_4(\mathbf{k},\mathbf{q};t) = 
\frac{1}{N} \left< \sum_{n,m} \hat{F}_n(\mathbf{k};t) \hat{F}_m(\mathbf{k};t)
e^{i \mathbf{q} \cdot [\mathbf{r}_n(0) - \mathbf{r}_m(0)]} \right>.
\end{equation}
Here $\hat{F}_n(\mathbf{k},t)$ is the microscopic self-intermediate scattering function,
$\hat{F}_n(\mathbf{k},t) = \cos[\mathbf{k} \cdot (\mathbf{r}_n(t) - \mathbf{r}_n(0))]$. 
Its ensemble average is the self-intermediate scattering function $F_s(k;t)$. 
A similar four-point structure factor was examined in Ref.~\cite{Flenner2007}.
 
The four-point structure factor $S_4(\mathbf{k},\mathbf{q};t)$ is sensitive to dynamics along the wave-vector 
$\mathbf{k}$. A slow spatial decay of correlations of the dynamics along $\mathbf{k}$ would be revealed in the
small $q$ values of $S_4(\mathbf{k},\mathbf{q};t)$. The spatial decay of correlations of the dynamics 
along the direction of 
the initial separation vector $\Delta \mathbf{r}_{nm}(0) = \mathbf{r}_n(0) - \mathbf{r}_m(0)$ are measured by
examination of $S_4(\mathbf{k},\mathbf{q};t)$ where $\mathbf{k}$ and $\mathbf{q}$ are parallel, and 
correlations of the dynamics along a direction perpendicular to 
$\Delta \mathbf{r}_{nm}(0)$ are measured by examination of $S_4(\mathbf{k},\mathbf{q};t)$ when $\mathbf{k}$
and $\mathbf{q}$ are perpendicular. 
We calculate $S_4(\mathbf{k},\mathbf{q};t)$ at a fixed angle $\theta$ 
between $\mathbf{k}$ and $\mathbf{q}$. 
We determine $\xi_4^\theta$ by fitting $S_4(\mathbf{k},\mathbf{q};t)$ using 
the same procedure described in Refs. \cite{Flenner2010,Flenner2011,Flenner2013}.
 
\begin{figure}
\includegraphics[width=3.2in]{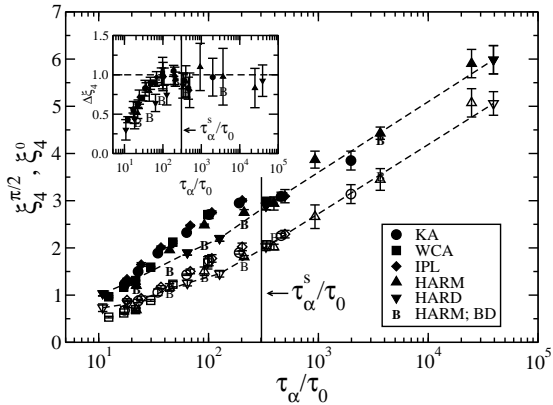}
\caption{\label{xione}The dynamic correlation length $\xi_4^0$ and $\xi_4^{\pi/2}$ as a 
function of the rescaled relaxation time $\tau_\alpha/\tau_0$. 
The dashed lines connect the data for the hard-sphere system. The points for the HARM system with
Brownian dynamics are marked with a $B$, and the error bars, which are about the size of the points,
are omitted for clarity. The inset shows the difference $\Delta \xi_4 = \xi_4^0 - \xi_4^{\pi/2}$ as a 
function of the rescaled relaxation time $\tau_\alpha/\tau_0$. The horizontal dashed line is $\Delta \xi_4 = 1.0$.}
\end{figure}

Shown in Fig.~\ref{xione} is $\xi_4^\theta$ for $\theta = 0$ and $\theta = \pi/2$ as a function
of $\tau_\alpha/\tau_0$ \cite{comment2}. The results for all the systems follow the same trend.
For the first 1.5 decades of slowing down correlations along the particles 
separation vector grows faster than correlations perpendicular to the
separation vector, and there is a small dynamics dependence in 
the growth of $\xi_4^\theta$, but there is
no dependence on the specifics of the interactions for this set of binary glass-formers. 
The similarity between the KA, WCA, and IPL systems indicates that there is no change 
in the shape of dynamically heterogeneous regions due to 
the presence of attractive interactions for this range of relaxation times.
This is qualitatively different from the results of Zhang \textit{et al.}\ \cite{Zhang2011}. We note that 
in the latter study clusters of fast particles were monitored whereas we examine correlations of slow particles.
In addition, Zhang \textit{et al.}\ examined dynamics in colloidal \emph{glasses} whereas we study 
equilibrium \emph{liquids} approaching the glass transition. 
  
We observe that for all systems the initial growth
of $\xi_4^0$ is faster than the initial growth of $\xi_4^{\pi/2}$, see inset to Fig.~\ref{xione} 
where we show $\Delta \xi_4 = \xi_4^0-\xi_4^{\pi/2}$.
The correlation length $\xi_4^0$ grows faster than $\xi_4^{\pi/2}$
until the difference between the two is around one particle diameter, then they grow at statistically
the same rate as a function of the rescaled relaxation time. The initial growth of 
$\Delta \xi_4$ depends slightly on the microscopic dynamics, but is independent of system.
For times exceeding $\tau_\alpha^s$, $\Delta \xi_4$ is
around one particle diameter and it is independent of the dynamics or the details of the interactions.
 
Having a larger $\xi_4^0$ than $\xi_4^{\pi/2}$ is suggestive of the string-like motion reported
in previous work \cite{Donati1998,Kim2000}. However, our work examines the slow particles, while the
string-like motion is observed for the fast particles. 
We leave a more detailed study of the
connection between our results and the string-like motion for future work.

In summary, we demonstrated universal behavior of the size and shape of dynamic heterogeneity 
for temperatures below $T_s$ (or volume fractions above $\phi_s$), \textit{i.e.}, 
below the temperature (above the volume fraction) where Stokes-Einstein violation
begins. We note that $T_s$ is below 
the the onset temperature of supercooling, $T_o$, thus below the temperature where dynamic
heterogeneity emerges. Thus, there is an intermediate temperature (volume fraction) regime where
the spatial extent of the dynamic heterogeneity is universal but its shape is dynamics-dependent.
We compared our results to predictions of different theories of glassy dynamics. In order to clearly 
differentiate between the RFOT theory and the facilitation approach we would need to extend the 
range of relaxation times by approximately two decades. This would also require simulating larger systems
and it is not feasible with our current computer resources.  

We note that our universal correlation between the dynamic correlation length and the relaxation time
parallels the correlation between the static point-to-set length and the relaxation time found by Hocky \textit{et al.}
Combining our results and those of Ref. \cite{Hocky2012} we could claim a correlation between the
dynamic correlation length and the static point-to-set length. However, there are two cautionary notes
regarding this possible relationship. First, we examined a significantly bigger range of slowing down whereas  
Hocky \textit{et al.} were restricted by the well-known difficulty of equilibrating systems in confinement.
Second, Hocky \textit{et al.}'s lengths were determined using the so-called spherical geometry
\cite{BerthierKob2012}. Charbonneau and Tarjus \cite{Charbonneau2013} 
used an alternative way to obtain the point-to-set lengths, the so-called random pinning geometry, 
and obtained static lengths that seem to be uncorrelated 
with dynamic correlation lengths. It is unclear whether the fundamental difference between Refs. \cite{Hocky2012} 
and \cite{Charbonneau2013},  
originates from different geometry and/or different systems used in these two studies. 
 
We gratefully acknowledge the support of NSF grant CHE 1213401.
This research utilized the CSU ISTeC Cray HPC System supported by NSF Grant CNS-0923386.

 \begin{figure*}
 \includegraphics[width=1.\textwidth]{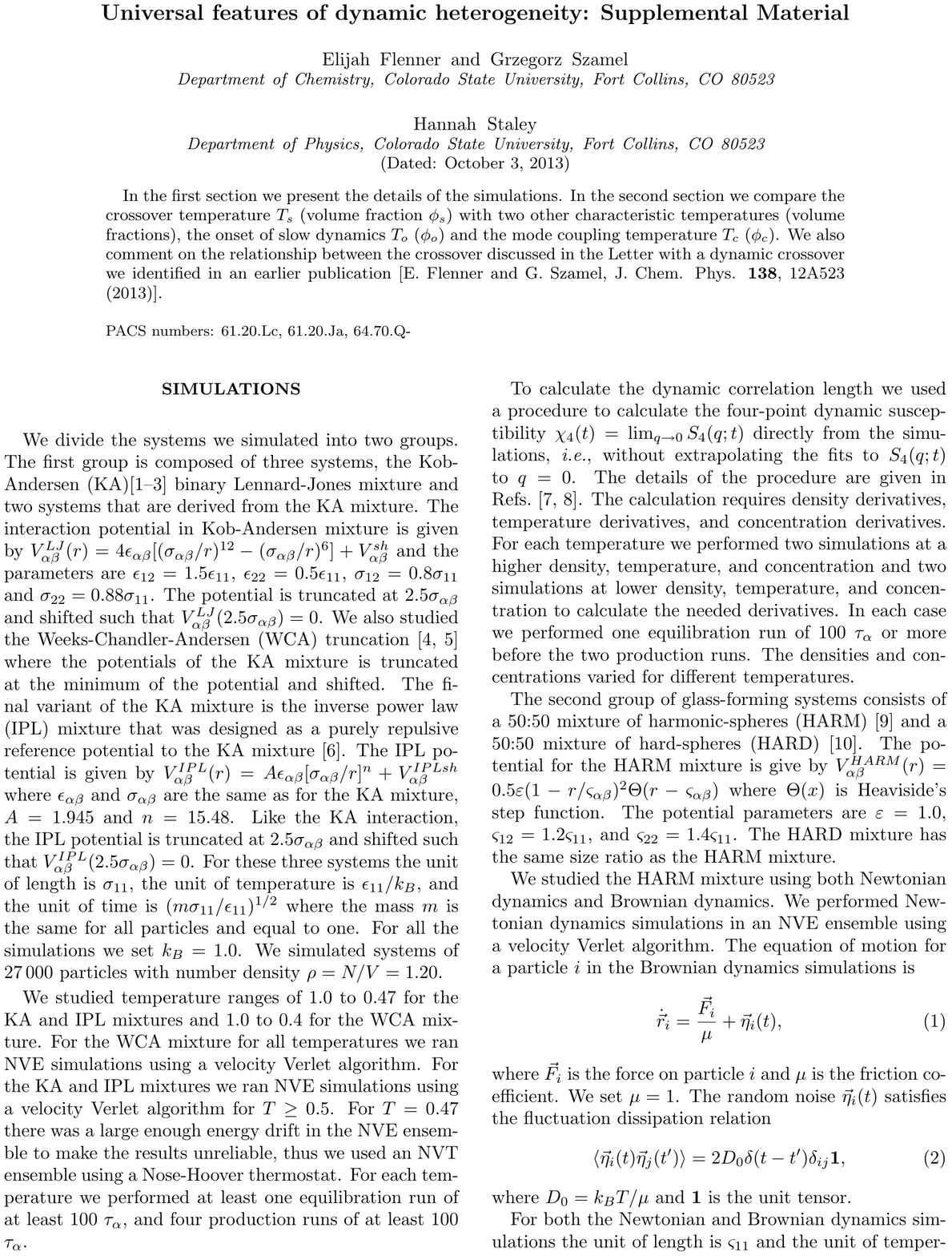}
 \end{figure*}
 \begin{figure*}
 \includegraphics[width=1.\textwidth]{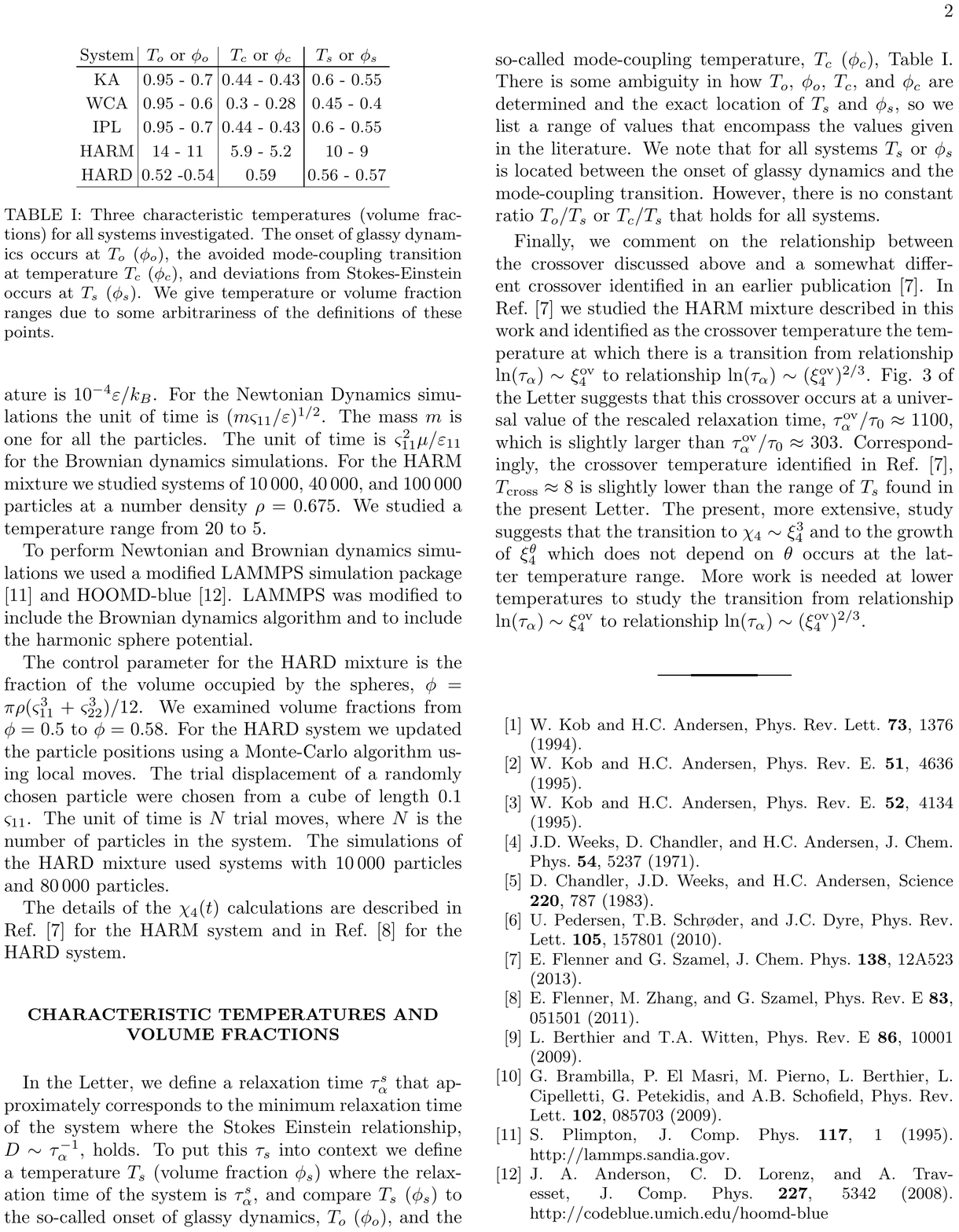}
 \end{figure*}
 
\end{document}